\documentclass[11pt,showpacs,preprintnumbers,superscriptaddress,amsmath,amssymb,nofootinbib]{revtex4}
\usepackage{graphicx}
\usepackage{dcolumn}
\usepackage{bm}
\usepackage{amssymb}
\usepackage{amsmath}
\usepackage{epsfig}    
\usepackage{color}
\usepackage{slashed}
\usepackage{hhline}
\usepackage{changes}

\def\be{\begin{equation}}
\def\ee{\end{equation}}
\def\Mat3#1#2#3#4#5#6#7#8#9{
\left(
\begin{array}{ccc}
#1 & #2 & #3 \\
#4 & #5 & #6 \\
#7 & #8 & #9 \\
\end{array}
\right) }
\newcommand{\bea}{\begin{eqnarray}}
\newcommand{\eea}{\end{eqnarray}}
\newcommand{\nn}{\nonumber}

\numberwithin{equation}{section}

\definechangesauthor[name=DW, color=red]{DW}

\begin{document}

\title{Muon $g-2$ in $U(1)_{\mu-\tau}$ Symmetric  Gauged Radiative Neutrino Mass Model}
\preprint{KIAS-P21026, APCTP Pre2021 - 016}

\author{Dong Woo Kang}
\email{dongwookang@kias.re.kr}
\affiliation{School of Physics, KIAS, Seoul 02455, Korea}
\author{Jongkuk Kim}
\email{jkkim@kias.re.kr}
\affiliation{School of Physics, KIAS, Seoul 02455, Korea}
\author{Hiroshi Okada}
\email{hiroshi.okada@apctp.org}
\affiliation{Asia Pacific Center for Theoretical Physics (APCTP) - Headquarters San 31, Hyoja-dong,
Nam-gu, Pohang 790-784, Korea}
\affiliation{Department of Physics, Pohang University of Science and Technology, Pohang 37673, Republic of Korea}

\begin{abstract}
{We explore muon anomalous magnetic moment (muon $g-2$) in a scotogenic neutrino model with a gauged lepton number symmetry $U(1)_{\mu-\tau}$. In this model, a dominant muon $g-2$ contribution comes not from an additional gauge sector but from a Yukawa sector.
In our numerical $\Delta \chi^2$ analysis, we show that our model is in favor of normal hierarchy with some features. We demonstrate one benchmark point, satisfying muon $g-2$ at the best fit value $25.1\times10^{-10}$.
}

\end{abstract}

\maketitle
\newpage

\section{Introduction}
Flavor dependent gauged $U(1)$ scenarios; especially $U(1)_{\mu-\tau}$~\cite{He:1990pn, He:1991qd}, are widely applied to explaining recent anomalies such as muon anomalous magnetic moment (muon $g-2$ or $\Delta a_\mu$)~\cite{Davier:2010nc, Davier:2017zfy, Davier:2019can, Borah:2021mri, Qi:2021rhh, Singirala:2021gok, Buras:2021btx, Zhou:2021vnf, Borah:2021jzu, Chen:2021vzk, Zu:2021odn, Huang:2021nkl, Patra:2016shz, Altmannshofer:2016oaq} and rare $B$ meson decays $b\to s \ell\bar\ell$~\cite{Crivellin:2015mga, Crivellin:2015lwa, Cen:2021iwv, Ko:2017yrd, Kumar:2020web} as beyond the Standard Model (SM), in addition to neutrino mass models~\cite{Baek:2015mna, Nomura:2018cle, Nomura:2018vfz, Asai:2018ocx, Asai:2017ryy, Lee:2017ekw}. \footnote{Flavor dependent gauge $U(1)$ scenarios can modify the lepton universality, since a new gauge boson $Z'$ does not universally couple to lepton sector. See \cite{Chun:2018ibr}. }
{{In fact, the longstanding muon $g-2$ anomaly was first discovered by BNL two decades ago and is recently confirmed by FNAL. Combined two results are given by}} ~\cite{Muong-2:2021ojo}  
\begin{equation}
\label{eq:amu_new}
\Delta a_\mu^{\rm new} = (25.1 \pm 5.9) \times 10^{-10} .
\end{equation}
The deviation from the SM expectation is 4.2$\sigma$ confidence level.
{One can typically address this deviation with the
	contribution from additional gauge boson with new neutral 
	gauge boson mass $m_{Z'}$ and coupling $g'$.
	}
Even though there are several constraints on this new gauge sector such as 
neutrino trident bound; $m_{Z'}/g'\lesssim 550$ GeV,
there exists allowed region~\cite{Altmannshofer:2014pba}. 

{In this paper, we propose {a scotogenic neutrino model with $U(1)_{\mu-\tau}$ gauge symmetry} that explains muon $g-2$ by the Yukawa sectors not by the new gauge sector.
	In order to get the sizable muon $g-2$, we need a diagram without chiral
	suppression and we introduce several exotic fermions and scalars.  
	These several exotic fermions can also play a role in radiative
	neutrino mass generation at one-loop level~\cite{Ma:2006km}.
	Heavier neutral fermions run inside the loop and the lightest one 
	could be a promising dark matter (DM) candidate.
}
Finally, we show several features of our model by demonstrating the numerical $\Delta \chi^2$ analysis.

This paper is organized as follows.
{	In Sec.~II, we present the model set up and how to generate 
	neutrino masses in normal and inverted hierarchy.
	In Sec.~III, we discuss constraints from lepton flavor violations (LFVs) and
	address the muon anomalous magnetic moment.
	We carry out numerical $\Delta \chi^2$ analysis and present the allowed 
	region satisfying the neutrino oscillation data, LFVs and muon $g-2$.
	Conclusions and discussions are given in Sec.~IV, briefly mentioning
	a possibility of DM candidate and how to explain the correct relic 
	density, satisfying direct detection bounds.
}

 \section{Model}
 \begin{widetext}
\begin{center} 
\begin{table}
\begin{footnotesize}
\begin{tabular}{|c||c|c|c|c|c|c|c|c|c|c|c|}\hline\hline  
 & \multicolumn{10}{c|}{Leptons} \\\hline
Fermions  ~&~ $L_{L_e}$ ~&~ $L_{L_\mu}$ ~&~ $L_{L_\tau}$ ~&~ $\ell_{R_e}$ ~&~ $\ell_{R_\mu}$ ~&~ $\ell_{R_\tau}$ ~&~ $N_{R_e}$ ~&~ $N_{R_\mu}$ ~&~ $N_{R_\tau}$~&~ $L'$~
\\\hline 
 $SU(2)_L$  & $\bm{2}$  & $\bm{2}$  & $\bm{2}$   & $\bm{1}$  & $\bm{1}$   & $\bm{1}$  & $\bm{1}$  & $\bm{1}$   & $\bm{1}$  & $\bm{2}$ \\\hline 
$U(1)_Y$  & $-\frac{1}{2}$ & $-\frac12$ & $-\frac12$  & $-1$ &  $-1$  &  $-1$  & $0$ &  $0$  &  $0$ & $-\frac12$ \\\hline
 $U(1)_{\mu-\tau}$ & $0$  & $1$ & $-1$ & $0$  & $1$   & $-1$ & $0$  & $1$   & $-1$ & $0$ \\\hline
$Z_2$  & $+$  & $+$ & $+$ & $+$ & $+$ & $+$& $-$ & $-$ & $-$& $-$ \\\hline
\end{tabular}
\caption{Field contents of fermions
and their charge assignments under $SU(2)_L\times U(1)_Y\times  U(1)_{\mu-\tau}\times Z_2$, where $SU(3)_C$ is singlet.}
\label{tab:1}
 \end{footnotesize}
\end{table}
\end{center}
\end{widetext}

\begin{table}[t]
\centering {\fontsize{10}{12}
\begin{tabular}{|c||c|c||c|c|c|}\hline\hline
&\multicolumn{2}{c||}{VEV$\neq 0$} & \multicolumn{2}{c|}{ VEV$= 0$ } \\\hline
  Bosons  &~ $H$  ~ &~ $\varphi$    ~ &~ $\eta$ ~&~ $\chi^-$ ~ \\\hline
$SU(2)_L$ & $\bm{2}$ & $\bm{1}$   & $\bm{2}$ & $\bm{1}$    \\\hline 
$U(1)_Y$ & $\frac12$ & $0$    & $\frac12$   & $-1$  \\\hline
 $U(1)_{\mu-\tau}$ & $0$  & $1$   & $0$ & $1$   \\\hline
$Z_2$ & $+$ & $+$    & $-$ & $-$ \\\hline
\end{tabular}%
} 
\caption{Field contents of bosons
and their charge assignments under $SU(2)_L\times U(1)_Y\times U(1)'\times Z_2$, where $SU(3)_C$ is singlet. }
\label{tab:2}
\end{table}

{In this section, we set up our model Lagrangian and focus on lepton sector and Higgs sector which are crucial for generating neutrino mass and muon g-2 contribution. And we address detail discussion for the neutrino mass matrix.}
	
\subsection{Lepton Lagrangian}
We introduce three right-handed neutral fermions $[N_{R_e},N_{R_\mu},N_{R_\tau}]$ with $[0,1,-1]$ charges under $U(1)_{\mu-\tau}$ symmetry and an isospin doublet vector-like lepton $L'\equiv [N',E']^T$ with $0$ under $U(1)_{\mu-\tau}$ symmetry.
In addition, we impose $Z_2$ odd for new fermions where even is assigned to the SM fermions.
The $Z_2$ symmetry plays a role in generating the active neutrino
mass matrix {not} at tree-level but one-loop level.
Furthermore, the lightest field with odd $Z_2$ can be a dark matter candidate.
Notice here that the $U(1)_{\mu-\tau}$ anomaly is independently canceled among the SM fermions or $N_R$.
In the bosonic sector, we introduce extra bosons with two isospin singlets $\varphi,\, \chi^-$ and a doublet $\eta$. These bosons respectively have $[1,1,0]$ charges under the $U(1)_{\mu-\tau}$ symmetry. 
{$\varphi$ is even under $Z_{2}$ and $\eta,\ \chi^-$ are odd under $Z_2$ and induce the active neutrino mass matrix together with $N_R$. The field contents and assignments for new fermions and the SM leptons are summarized in Tables~\ref{tab:1}.}
Each of $\varphi$ and the SM Higgs denoted by $H$ has non-zero vacuum expectation values that are symbolized by $\langle\varphi\rangle\equiv v_\varphi/\sqrt2$ and  $\langle H\rangle\equiv [0,v_H/\sqrt2]^T$, after spontaneous symmetry breaking of $U(1)_{\mu-\tau}$  and electroweak symmetry.
The bosonic field contents and their assignments are summarized in Tables~\ref{tab:2}.

Under these symmetries, our renormalizable Lagrangian is given by
\begin{align}
-\mathcal{L}_Y & = 
M_{ee} \overline{N_{R_e}^{c}} N_{R_e} + M_{\mu\tau} (\overline{N_{R_\mu}^c} N_{R_\tau} + \overline{N_{R_\tau}^{c}} N_{R_\mu})  
+ M_{L'} \overline{L'_L} L'_R +\text{h.c.} \notag\\
&+y_e \overline{L_{L_e}} H e_R +y_\mu \overline{L_{L_\mu}} H \mu_R + y_\tau \overline{L_{L_\tau}} H \tau_R+ {\rm h.c.}\notag\\
&+h_{e\mu}^{}(\overline{N_{R_e}^{c}} N_{R_\mu}+\overline{N_{R_\mu}^{c}}N_{R_e}) \varphi^* 
+ h_{e\tau}^{}(\overline{N_{R_e}^c } N_{R_\tau} + \overline{N_{R_\tau}^{c}} N_{R_e} )\varphi + {\rm h.c.} \notag\\
&+f_e \overline{L_{L_e}}(i\sigma_2)\eta^* N_{R_e}  
 + f_\mu \overline{L_{L_\mu}} (i\sigma_2)\eta^* N_{R_\mu} 
 + f_\tau \overline{L_{L_\tau}} (i\sigma_2)\eta^* N_{R_\tau}
 \notag\\
&+g_e \overline{\ell_{R_e}} N_{R_\mu}^c \chi^-  
 + g_\mu  \overline{\ell_{R_\mu}} N_{R_e}^c \chi^-
  \notag\\
&+ h \overline{L_{L_\mu}} L'^c_L \chi^-  
 + y'  \overline{L'_L}(i\sigma_2) H^* N_{R_e}
  + y''  \overline{L'^c_R} H N_{R_e}
+  {\rm h.c.} ,
\label{yukawa}
\end{align}
where $\sigma_2$ is the second Pauli matrix and  the charged-lepton mass matrix is diagonal due to the $\mu-\tau$ symmetry; $\mathcal{M}_\ell = \frac{v_H}{\sqrt{2}}\text{diag}(|y_e|,|y_\mu|,|y_\tau|)\equiv (m_e,m_\mu,m_\tau)$ after the phase redefinition. Therefore, the neutrino oscillation data is induced via neutrino sector.
\subsection{Higgs sector}
Our Higgs potential is also given by
\begin{align}
 {\cal V} &= {\cal V}_{2}^{tri} + {\cal V}_4^{tri}
 + \lambda_0(H^\dag\eta)^2 +\lambda'_0\varphi^* (H^T i\sigma_2 \eta)\chi^-
 + {\rm h.c.}
,\label{Eq:pot}
\end{align}
where we define $\eta\equiv [\eta^+,(\eta_R+i\eta_I)/\sqrt2]^T$, $H\equiv [h^+,(v_H+h_0+iz_0)/\sqrt2]^T$, $\varphi\equiv(v_\varphi+\varphi_R+i z_\varphi)/\sqrt2$,
and ${\cal V}_{2}^{tri}$ and ${\cal V}_4^{tri}$ are respectively trivial quadratic and quartic terms of the Higgs potential;
$
{\cal V}_{2}^{tri}= \sum_{\phi=H,\varphi,\eta,\chi^-} \mu_{\phi}^2|\phi|^2,\quad
{\cal V}_{4}^{tri}= \sum_{\phi'\le\phi}^{H,\varphi,\eta,\chi^-} \lambda_{\phi\phi'} |\phi|^2|\phi'|^2+\lambda'_{H\eta}|H^\dag \eta|^2.\label{Eq:pot-tri}
$
Notice here that $h^+$, $z_0$, and $z_\varphi$ are respectively absorbed by the longitudinal degrees of freedom in gauge sectors. 
Consequently, we have massive gauge bosons $W^\pm, Z$ in the SM and $Z'$ in the $U(1)_{\mu-\tau}$ gauge symmetry.
{The $\lambda_0$ term plays an important role in generating the non-vanishing neutrino mass matrix.
In our model, the neutrino mass matrix is proportional to the mass-squared difference between $\eta_R$ and $\eta_I$}; $m_R^2-m_I^2= \lambda_0 v_H^2$, where $m_{R,I}$ is the mass eigenstate of $\eta_{R,I}$~\cite{Ma:2006km}.
Even though there is mixing between $\chi^\pm$ and $\eta^\pm$ from $\lambda'_0$,
we {suppose} that the mixing is negligibly tiny.~\footnote{If we consider the large mixing, we have a contribution to muon $g-2$ without chiral suppression, which might lead to large muon $g-2$. 
However, it cannot be large enough. In fact, we need large mass hierarchy between them, but it is forbidden by oblique parameters~\cite{Peskin:1991sw}. Satisfying these conditions, we have found $\Delta a_\mu \approx 10^{-13}$ at most in our numerical estimation.}

\subsection{Neutral fermion mass matrices}
After the phase redefinition of the neutral fermions, 
the mass matrix in basis of $[N_{R_e},N_{R_\mu},N_{R_\tau},N'^c_L, N'_R]$ is found as follows:
\begin{align}
\mathcal{M}_N  = 
\begin{pmatrix}
M_{ee} &M_{e\mu} & M_{e\tau} & m' & m'' e^{i\zeta}\\ 
M_{e\mu}  & 0 & M_{\mu\tau} e^{i\xi} & 0 & 0 \\
M_{e\tau} & M_{\mu\tau} e^{i\xi} & 0 & 0 & 0 \\
 m' &0 & 0 & 0 & M_{L'} \\
 m'' e^{i\zeta} &0 & 0 & M_{L'} & 0 \\
\end{pmatrix},   \label{massmat}
\end{align}
where $M_{e\mu}\equiv \frac{v_\varphi}{\sqrt{2}}|h_{e\mu}|$, $M_{e\tau}\equiv \frac{v_\varphi}{\sqrt{2}}|h_{e\tau}|$
, $m' \equiv \frac{v_H}{\sqrt{2}}|y'|$,  $m'' \equiv \frac{v_H}{\sqrt{2}}|y''|$ are real mass parameters, while $\xi,\ \zeta$ are physical phases.
{The mass matrix $\mathcal{M}_N$ is then diagonalized by introducing a 	unitary matrix $V$. This matrix satisfies }
\begin{align}
 V^T \mathcal{M}_N V
 \equiv \text{diag}(M_{1},M_{2},M_{3},M_{4},M_{5}). 
\end{align}
Here, the mass eigenstate $\psi_R$ is defined by $N_{R_i}=\sum_{k=1}^5 V_{i k} \psi_{R_k}$, and its mass eigenvalue is defined by $M_k(k=1,2,3,4,5)$.
Then, the valid Lagrangian is rewritten in terms of mass eigenstates as follows
\footnote{{Notice here that $h \bar \nu_{\mu L} E'^C_L \chi^-$ does not contribute to the neutrino mass matrix, since $E'^C_L$ cannot propagate in the loop. }}:
\begin{align}
-{\cal L} &= \bar\nu_{L_a} F_{ak} \psi_{R_k}(\eta_R-i \eta_I)
- \sqrt{2} \bar\ell_{L_a} F_{ak} \psi_{R_k}\eta^- 
+ \bar\ell_{R_{a'}} G_{a' k} \psi_{R_k}^C \chi^-
+ H_{4k} \bar\mu_L \psi_{R_k} \chi^-
+{\rm h.c.},\label{eq:lgrg}\\
F_{ia} &
=\frac1{\sqrt2}
\sum_{j=1,2,3}\begin{pmatrix}
f_e & 0 & 0 \\ 
0  & f_\mu &0  \\
0&0 & f_\tau   \\
\end{pmatrix}_{ij} V_{j a},\quad 
G_{ia} 
=
\sum_{j=1,2,3}\begin{pmatrix}
0 & g_e & 0 \\ 
g_\mu  & 0&0  \\
\end{pmatrix}_{ij} V^*_{ja},\quad
H_{4k}\equiv h \sum_{k=1}^5 V_{4k},
\end{align}
where $F$ is three by five matrix, and $G$ is two by five matrix, therefore $a'$ runs over $e,\mu$.
The first term of Eq.~(\ref{eq:lgrg}) contributes to the neutrino mass matrix, while the other terms induce LFVs, and muon $g-2$ as can be seen below.
The active neutrino mass matrix is given by~\cite{Ma:2006km} 
\begin{align}
(m_{\nu})_{ij}
&=\sum_{k=1}^{5} 
\frac{F_{ik} M_{k} F_{kj}^T}{2 (4\pi)^2}
\left[ \frac{m^2_{R}}{m^2_{R}-M^2_k} \ln \frac{m^2_{R}}{M^2_k}-
\frac{m^2_{I}}{m^2_{I}-M^2_k} \ln \frac{m^2_{I}}{M^2_k}  \right]~\nn\\
&\simeq \frac{\lambda_0 v^2_H}{(4\pi)^2}\sum_{k=1}^{5} 
 \frac{F_{ik} M_k F_{kj}^T}{m_0^2-M^2_k}
\left[ 1-\frac{M^2_k}{m^2_0-M^2_k} \ln \frac{m^2_0}{M^2_k}  \right] ~,
\end{align}
where we assume to be $\lambda_0 v_H^2 = m^2_R-m^2_I \ll m_0^2\equiv (m_R^2+m_I^2)/2$ in the second line, 
and $m_\nu$ is diagonalzied by a unitary matrix $U_{\rm PMNS}$~\cite{Maki:1962mu}; $D_\nu\equiv U_{\rm PMNS}^T m_\nu U_{\rm PMNS}$.
Here, we define dimensionless neutrino mass matrix as $m_\nu\equiv (\lambda_0 v_H) \tilde m_\nu\equiv\kappa \tilde m_\nu$.
Since $\kappa$ does not depend on the flavor structure, we rewrite this diagonalization in terms of dimensionless form $(\tilde D_{\nu_1},\tilde D_{\nu_2},\tilde D_{\nu_3}) \equiv U_{\rm PMNS}^T \tilde m_\nu U_{\rm PMNS}$.
Thus, we fix $\kappa$ by
\begin{align}
({\rm NH}):\  \kappa^2= \frac{|\Delta m_{\rm atm}^2|}{\tilde D_{\nu_3}^2-\tilde D_{\nu_1}^2},
\quad
({\rm IH}):\  \kappa^2= \frac{|\Delta m_{\rm atm}^2|}{\tilde D_{\nu_2}^2-\tilde D_{\nu_3}^2},
 \end{align}
where $\Delta m_{\rm atm}^2$ is the atmospheric neutrino mass-squared difference.  
{Here, NH and IH stand for the normal hierarchy and the inverted hierarchy, respectively.} 
Subsequently, the solar neutrino mass-squared difference is depicted in terms of $\kappa$ as
follows:
\begin{align}
\Delta m_{\rm sol}^2= {\kappa^2}({\tilde D_{\nu_2}^2-\tilde D_{\nu_1}^2}).
 \end{align}
This should be within the experimental value. 
 The neutrinoless double beta decay is also given by 
\begin{align}
\langle m_{ee}\rangle=\kappa\left|\tilde D_{\nu_1} \cos^2\theta_{12} \cos^2\theta_{13}+\tilde D_{\nu_2} \sin^2\theta_{12} \cos^2\theta_{13}e^{i\alpha_{2}}+\tilde D_{\nu_3} \sin^2\theta_{13}e^{i(\alpha_{3}-2\delta_{CP})}\right|,
\end{align}
which  may be able to observed by KamLAND-Zen in future~\cite{KamLAND-Zen:2016pfg}. 

\section{ Results} 
\subsection{Lepton flavor violations} \label{lfv-lu}
{	Because of the flavor dependence in Eq.(\ref{eq:lgrg}), 
	we have lepton flavor violation decay processes 
	for example, $\mu\to e\gamma$ or $\bar\mu\to\bar e\gamma$.
	These decay channels are originated from the terms of 
	$G H^\dag$ or $H G^\dag$.
}\footnote{In our numerical analysis, we also have considered non-dominant contributions of $G G^\dag$ and $F F^\dag$.} 
The corresponding branching ratio is given as follows~\cite{Lindner:2016bgg, CarcamoHernandez:2019ydc, Baek:2016kud}
\begin{align}
{\rm BR}(\bar\mu\to \bar e\gamma)
&\simeq
\frac{48\pi^3 \alpha_{\rm em}  }{(4\pi)^4m_{\ell_i}^2 G_F^2} |H_{4k} M_k G^\dag_{k1} F(M_k,m_\chi)|^2,\\
{\rm BR}(\mu\to e\gamma)
&\simeq
\frac{48\pi^3 \alpha_{\rm em}  }{(4\pi)^4m_{\ell_i}^2 G_F^2} |G_{1k} M_k H^\dag_{k4} F(M_k,m_\chi)|^2,\\
F(m_1,m_2)& =\int[dx]_3 \frac{y}{(x^2-x) m^2_\mu + x m_1^2 + (y+x) m_2^2},
\label{eq:damu1}
\end{align}
where the fine structure constant $\alpha_{\rm em} \simeq 1/135$ at muon mass energy scale~\cite{TheMEG:2016wtm}, and the Fermi constant $G_F \simeq 1.17\times 10^{-5}$ GeV$^{-2}$. 
The current experimental upper bound at 90\% C.L. are~\cite{TheMEG:2016wtm, Adam:2013mnn}
\begin{align}
{\rm BR}(\mu\to e\gamma) = {\rm BR}(\bar\mu\to \bar e\gamma) < 4.2\times10^{-13} .
\end{align}

\subsection{Muon anomalous magnetic moment}
{In our model, we can achieve the required muon $g-2$ contribution through the same interaction with LFVs. The muon $g-2$ contribution is given as  } 
 \begin{align}
  & \Delta a_\mu \approx - \frac{m_\mu}{(4\pi)^2}{\rm Re}[(H_{4k} M_k G^\dag_{k2} + G_{2k} M_k H^\dag_{k4}) F(M_k,m_\chi)]
  =
  - \frac{2m_\mu}{(4\pi)^2}{\rm Re}[H_{4k} M_k G^\dag_{k2}] F(M_k,m_\chi)
  \label{amu1L}.
 \end{align}
Recent experimental estimation~\cite{Hanneke:2008tm} of muon $g-2$ indicates the following value at $4.2\sigma$~\cite{Muong-2:2021ojo}:
 \begin{align}
   \Delta a_\mu = a_\mu^{\rm{EXP}}-a_\mu^{\rm{SM}} = (25.1\pm 5.9)\times 10^{-10}.
   \label{eq:yeg2}
 \end{align}
%
We fix the best fit (BF) value; $ \Delta a_\mu= 25.1\times 10^{-10}$, in our numerical analysis below.

\subsection{Numerical analysis \label{sec:NA}}
{We show our numerical $\Delta \chi^2$ analysis to satisfy the neutrino oscillation data, LFV ($\mu\to e\gamma$ only in our case) as well as muon $g-2$. Here we fix $\Delta a_\mu$ as BF value $ 25.1\times 10^{-10}$, where we adopt five known observables  $\sin^2\theta_{12},\ \sin^2\theta_{23},\ \sin^2\theta_{13},\ \Delta m^2_{\rm atm},\ \Delta m^2_{\rm sol}$ in Nufit 5.0~\cite{Esteban:2020cvm} at $3\sigma$ confidence level. 
Notice that we do not include the Dirac CP phase in $\Delta \chi^2$ analysis because of big ambiguity at $3\sigma$ interval. Furthermore, we employ Gaussian approximations for charged-lepton masses.
}
{However since we have found the allowed region within $4.59\lesssim \sqrt{\Delta \chi^2}$ in case of IH, it does not satisfy the sizable muon $g-2$ whose maximum order is $10^{-12}$ in our model.
Thus, we focus on NH only.}
%
At first, we fix the range of our input parameters as follows:
\begin{align}
&\{ f_e, f_\mu, f_\tau \} \in [10^{-5},10],\ \{ h,|g_\mu| \} \in [1,10],\ \{ \xi,\zeta\} \in [0,\pi],\quad
 \{ m',m'' \} \in [1, 617]\ {\rm GeV},
\\
& \{ M_{ee}, M_{e\mu}, M_{e\tau}, M_{\mu\tau}\} \in [10,10^6]\ {\rm GeV},
\quad\{  m_{\eta^+}, M_{L'}, m_{\chi^\pm}, m_0 \} \in [10^2,10^6]\ {\rm GeV},
\end{align}
where $0.9\times m_0\le m_{\eta^\pm}\le 1.2\times m_0$ to evade the bound on oblique parameters, and $g_e=0$ to suppress the LFVs in our work.
\footnote{{In the case of $g_e=0$, both BR$(\mu\to e\gamma)$ and muon $g-2$ are proportional to Re[$h g_\mu^*$] that does not depend on the flavor structure. Thus, we can take any values for $h$ and $g_\mu$ in our numerical ranges as far as Re[$h g_\mu^*$] does not change. In our benchmark points in Table~\ref{bp-tab_nh}, we set $h$ and $g_\mu$ as one of examples so as to be Re[$h g_\mu^*$]$\sim$1.52.}}
The upper bounds on $m',\ m''$ come from $|y^{'('')}| v_H/\sqrt{2}$ taking the limit of perturbation $|y^{'('')}|=\sqrt{4\pi}$ in addition to $\{ F,\ G,\ H\} \in \sqrt{4\pi}$.

\begin{figure}[htbp]
  \includegraphics[width=54mm]{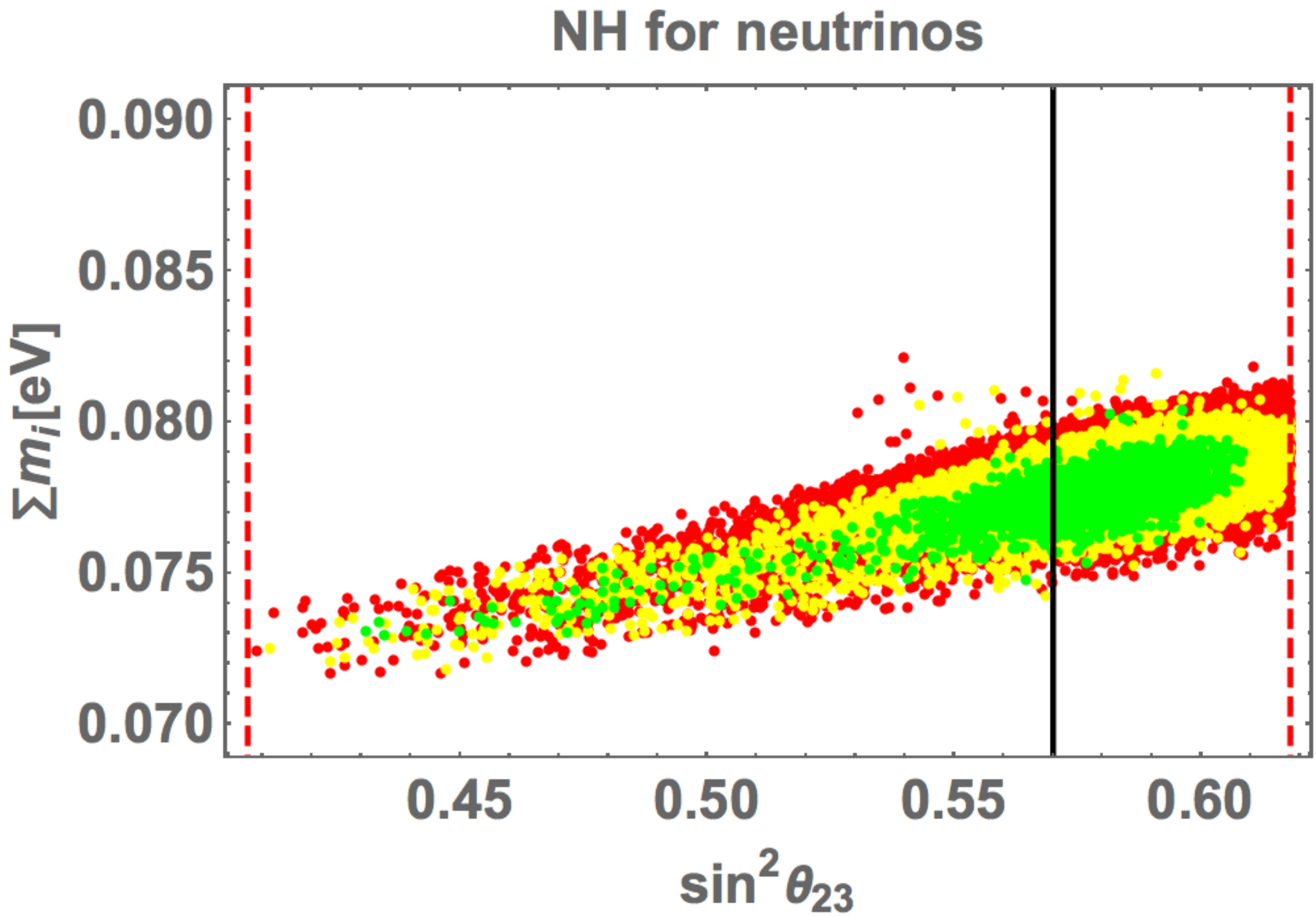}
    \includegraphics[width=54mm]{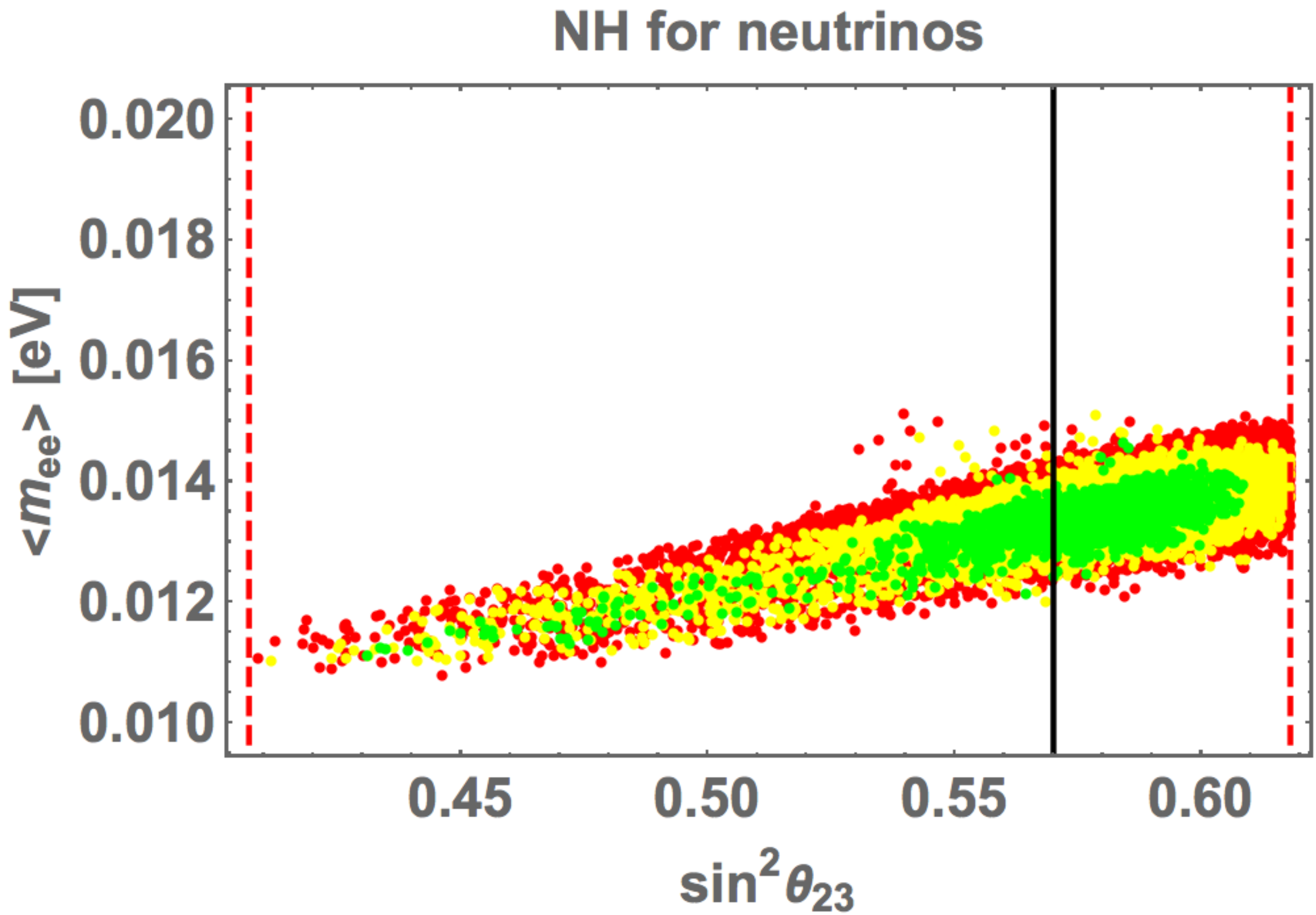}
  \includegraphics[width=54mm]{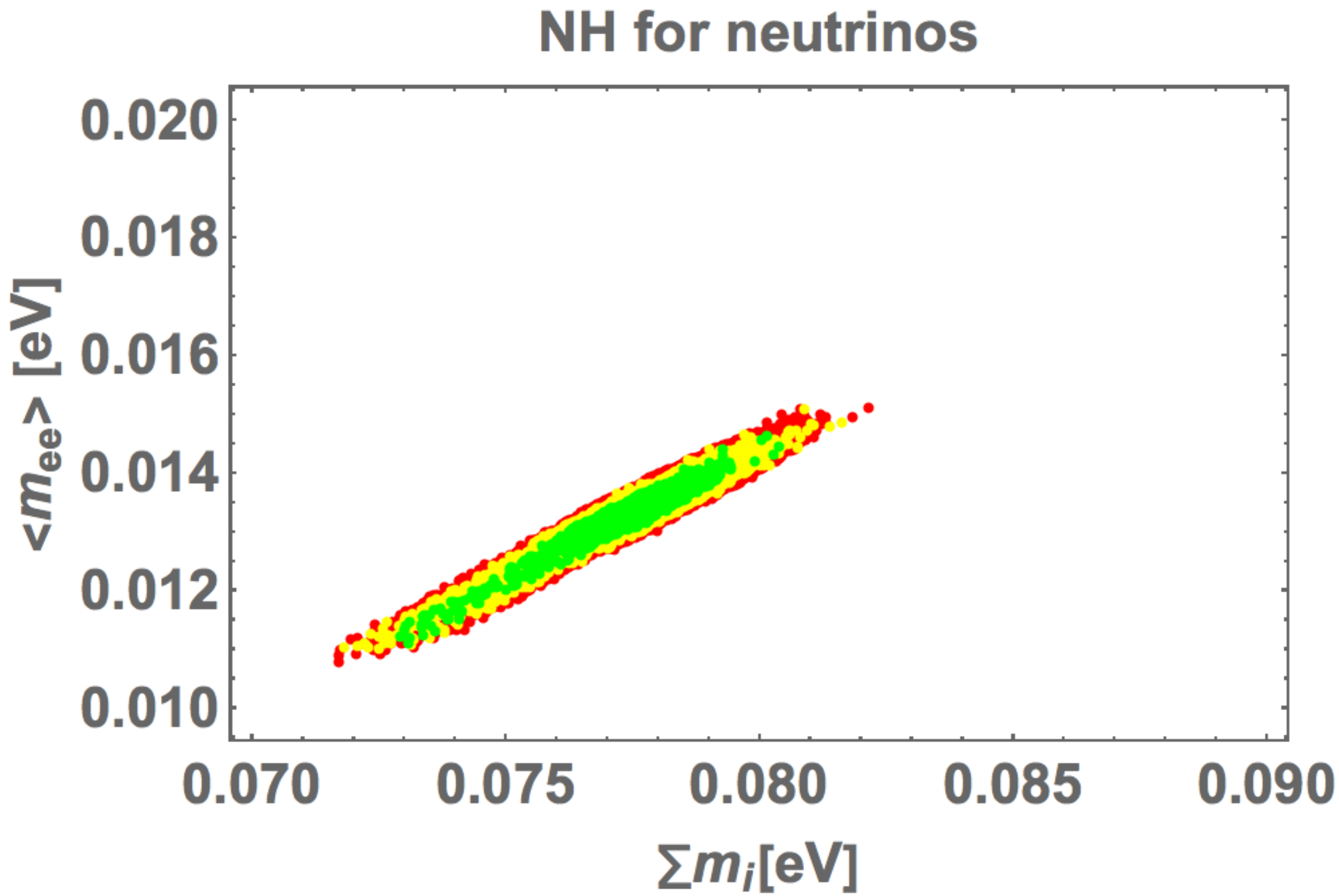}
  \caption{The scatter plots of sum of the neutrino masses $\sum m_i$ in terms of $\sin^2\theta_{23}$ (left), neutrinoless double beta decay $\langle m_{ee}\rangle$ in terms of $\sin^2\theta_{23}$ (center), and $\langle m_{ee}\rangle$ in terms of $\sum m_i$ (right).
Each of the color plots of green, yellow, and red represents  the region at $\sigma$ of $0$--$1$, $1$--$2$, $2$--$3$, respectively. The each vertical line of black-solid and  red-dotted represents the best fit and 3$\sigma$ interval. }
  \label{fig:s23-sum_NH}
\end{figure}
In Fig.~\ref{fig:s23-sum_NH}, we show the scatter plots of sum of neutrino masses $\sum m_i$ in terms of $\sin^2\theta_{23}$ (left), neutrinoless double beta decay $\langle m_{ee}\rangle$ in terms of $\sin^2\theta_{23}$ (center), and $\langle m_{ee}\rangle$ in terms of $\sum m_i$ (right).
Here, each of the color plots of green, yellow, and red represents the region at $\sigma$ of $0$--$1$, $1$--$2$, $2$--$3$, respectively. 
{After the $\Delta\chi^{2}$ analysis, as shown in the figure, we find that}
$\sum m_i$ is allowed in the range of $[72.9-80.4]$ [meV] at $1\sigma$ interval,  $[71.8-81.6]$ [meV] at $2\sigma$ interval,  $[71.7-82.1]$ [meV] at $3\sigma$ interval.
$\langle m_{ee}\rangle$ is allowed in the range of $[11.1-14.6]$ [meV] at $1\sigma$ interval, $[11.0-15.1]$ [meV] at $2\sigma$ interval, and $[10.8-15.1]$ [meV] at $3\sigma$ interval.

\begin{table}[h]
	\centering
	\begin{tabular}{|c|c|} \hline 
			\rule[14pt]{0pt}{0pt}
 		&  NH  \\  \hline
			\rule[14pt]{0pt}{0pt}
		$h$&   $ -{1.005}$   \\ \hline 
		\rule[14pt]{0pt}{0pt}
		$g_\mu$ & ${1.514+0.628 i}$  \\ \hline
		\rule[14pt]{0pt}{0pt}%
		$[f_e, f_\mu,f_\tau]$ & $[14.5,\ -7.78,\ -17.7]\times10^{-5}$     \\ \hline
		\rule[14pt]{0pt}{0pt}
		$[\xi, \zeta]$ & $[328^\circ, 209^\circ]$     \\ \hline
		\rule[14pt]{0pt}{0pt}
		$[m_0,m_{\chi^\pm}, m_{\eta^\pm},]$ & $[123,\ 165,\ 102]$\ {\rm GeV}    \\ \hline
		\rule[14pt]{0pt}{0pt}
		$\begin{pmatrix}
M_{ee} &M_{e\mu} & M_{e\tau} & m' & m'' \\ 
M_{e\mu}  & 0 & M_{\mu\tau}  & 0 & 0 \\
M_{e\tau} & M_{\mu\tau}   & 0 & 0 & 0 \\
 m' &0 & 0 & 0 & M_{L'} \\
 m''  &0 & 0 & M_{L'} & 0 \\
\end{pmatrix}$  & 
 $\begin{pmatrix}
40.2 & 2.15\times10^4 & 8.12\times10^4 &587 &-88.5\\ 
2.15\times10^4 & 0 & 8.34\times10^3  & 0 & 0 \\
8.12\times10^4 &8.34\times10^3   & 0 & 0 & 0 \\
587 &0 & 0 & 0 & 486\\
-88.5 &0 & 0 & 486 & 0 \\
\end{pmatrix}${\rm GeV}     \\ \hline
		\rule[14pt]{0pt}{0pt}
		${\rm BR}(\mu\to e\gamma)$  &  $4.11\times10^{-22} $   \\ \hline
		\rule[14pt]{0pt}{0pt}
		$\Delta m^2_{\rm atm}$  &  $2.52\times10^{-3} {\rm eV}^2$   \\ \hline
		\rule[14pt]{0pt}{0pt}
		$\Delta m^2_{\rm sol}$  &  $7.44\times10^{-5} {\rm eV}^2$    \\ \hline
		\rule[14pt]{0pt}{0pt}
		$\sin^2\theta_{12}$ & $ 0.307$  \\ \hline
		\rule[14pt]{0pt}{0pt}
		$\sin^2\theta_{23}$ &  $ 0.573$  \\ \hline
		\rule[14pt]{0pt}{0pt}
		$\sin^2\theta_{13}$ &  $ 0.0221$  \\ \hline
		\rule[14pt]{0pt}{0pt}
		$[\delta_{CP}^\ell,\ \alpha_{21},\,\alpha_{31}]$ &  $[0.191^\circ,\, 360^\circ,\, 0.405^\circ]$   \\ \hline
		\rule[14pt]{0pt}{0pt}
		$\sum m_i$ &  $77.5$\,meV     \\ \hline
		\rule[14pt]{0pt}{0pt}
		$\langle m_{ee} \rangle$ &  $13.3$\,meV     \\ \hline
		\rule[14pt]{0pt}{0pt}
		$\kappa$ &  $5.33${\rm GeV}   \\ \hline
		\rule[14pt]{0pt}{0pt}
		$\sqrt{\Delta\chi^2}$ &  $0.391$   \\ \hline
		\hline
	\end{tabular}
	\caption{Numerical benchmark points of our input parameters and observables in NH, satisfying muon $g-2$ at the best fit value $25.1\times10^{-10}$. Here, the NH is taken such that $\sqrt{\Delta \chi^2}$ should be minimum.}
	\label{bp-tab_nh}
\end{table}
%
Finally, we show a benchmark point for NH in Table~\ref{bp-tab_nh}, where it is taken such that $\Delta \chi^2$ is minimum in our model. Notice here that all the phases $\delta_{CP}$, $\alpha_{21},\ \alpha_{31}$ are very close to be zero through our $\sqrt{\Delta \chi^2}$ analysis.

\section{Conclusions and Discussions}
{We have proposed a gauged $U(1)_{\mu-\tau}$ neutrino mass model which  can address the muon $g-2$ through the Yukawa sector without relying on the new gauge sector.}
The neutrino mass matrix is induced at one-loop level and participated in five neutral fermions run inside the loop.
%
Through our numerical $\Delta \chi^2$ analysis, we have found that NH is in favor of our model.
{The best fit value of $\Delta a_\mu$ is mainly obtained by Re[$hg^*_\mu$] in case of $g_e=0$ for NH. 
IH does not satisfy the sizable muon $g-2$, but we have allowed region at $4.59\lesssim\sqrt{\Delta \chi^2}$ in neutrino oscillation data.}
Several unique features are listed below:
\begin{enumerate}
\item   $\sum m_i$ is allowed in the range of $[72.9-80.4]$ [meV] at $1\sigma$ interval,  $[71.8-81.6]$ [meV] at $2\sigma$ interval,  $[71.7-82.1]$ [meV] at $3\sigma$ interval.
\item  $\langle m_{ee}\rangle$ is allowed in the range of $[11.1-14.6]$ [meV] at $1\sigma$ interval, $[11.0-15.1]$ [meV] at $2\sigma$ interval, and $[10.8-15.1]$ [meV] at $3\sigma$ interval.
%
%
\item All the phases $\delta_{CP}$, $\alpha_{21},\ \alpha_{31}$ are localized in vicinity of zero through our ${\Delta \chi^2}$ analysis.
\end{enumerate}

Before closing our discussion, we briefly mention a DM candidate.
Basically, we have two candidates; bosonic and fermionic one.\\
{\it In case of bosonic DM candidate}, the lightest particle of $\eta_R$ and $\eta_I$ is the one.
{\it In case of fermionic DM candidate}, $\psi_{R_1}$ is the one.
In the case of fermionic DM candidate, we do not need to consider the bound on direct detection, because it does not interact with quark sector directly.
The dominant cross section arises from $G$ and $H$ terms that also appear in muon $g-2$ and have s-wave dominant.
But this contribution is at most ${\cal O}(5\times10^{-12})$ GeV$^{-2}$ that is too small to resolve the correct relic density.
Thus, {our promising DM candidate is bosonic.}
Here, let us suppose $\eta_R$ to be DM,
and we simply neglect any interactions coming from Higgs potential in order to evade any bounds from direct detection experiments.~\footnote{This assumption would be reasonable since these couplings are highly suppressed by bound on direct detection searches. The coupling among DM and Higgses are of the order $10^{-3}$~\cite{Kanemura:2010sh}.}
Moreover, we assume that the mass difference between $\eta_R$ and $\eta_I$ is more than 200 keV in order to evade the
inelastic direct detection bounds via $Z$ boson portal~\cite{Barbieri:2006dq}.
Similar to the result of fermionic case, we cannot rely on any contributions from Yukawa couplings because these constributions are too tiny to explain the relic density.
Thus, we have to make the use of kinetic interactions from the SM.
In this situation, nature of the DM  is seriously analyzed by e.g. Ref.~\cite{Hambye:2009pw}.
The solutions are uniquely found at the points of half of the Higgs mass; $\sim63$ GeV and 534 GeV
where{ coannihilation processes such as $\eta_R\eta_I\to Z\to \bar ff$ are taken in consideration.

\vspace{0.5cm}
\hspace{0.2cm} 

{\it Acknowledgments}
The work is supported in part by KIAS Individual Grants, Grant No. PG074202 (JK) and No. PG076201 (DK) at Korea Institute for Advanced Study.
This research was supported by an appointment to the JRG Program at the APCTP through the Science and Technology Promotion Fund and Lottery Fund of the Korean Government. This was also supported by the Korean Local Governments - Gyeongsangbuk-do Province and Pohang City (H.O.). 
H.O.~is sincerely grateful for all the KIAS members.

\end{document}